\documentclass[a4paper,fleqn,usenatbib]{mnras}  
  
\usepackage{mathptmx}
\usepackage{txfonts}  
\usepackage{bold-extra}  
\usepackage[T1]{fontenc}
\usepackage[T1]{fontenc}  
\usepackage{ae,aecompl}  
  
\usepackage{graphicx}	
  
\newcommand{\TS}{RGB~2243\, }  
\newcommand{\ts}{RGB~2243}   
\newcommand{\kms}{km~s$^{-1}$}

\newcommand{\ergscm}{erg~s$^{-1}$cm$^{-2}$}  
  
\title[Constraining the Redshift of the BL Lac  RGB~J2243+203]{New GTC Spectroscopic Data  
and a Statistical Study to Better Constrain the Redshift of the BL Lac  RGB~J2243+203}  
  
\author[Rosa Gonz\'alez,  et al.]{  
D. Rosa Gonz\'alez,$^{1}$\thanks{E-mail: danrosa@inaoep.mx} H. Muriel,$^{2,3}$  
Y. D. Mayya,$^{1}$ I. Aretxaga,$^{1}$ J. Becerra Gonz\'alez,$^{4,5}$\newauthor  
A. Carrami\~nana,$^{1}$  J. M\'endez-Abreu,$^{4,5}$  
O. Vega,$^{1}$ E. Terlevich,$^{1}$ S. Couti\~no de Le\'on,$^{1}$ \newauthor    
A. Furniss,$^{6}$ A. L. Longinotti,$^{1}$ R. J. Terlevich,$^{1}$ A. C. Pichel,$^{7}$ A. C. Rovero,$^{7}$ and  
\newauthor C. Donzelli $^{2,3}$  
  \\  
$^{1}$ Instituto Nacional de Astrof\'\i sica, Optica y Electr\'onica, Tonantzintla, 72840 Puebla, Mexico\\  
  $^{2}$  Instituto de Astronom\'\i a Te\'orica y Experimental (IATE),  
  CONICET-Universidad Nacional de C\'ordoba, C\'ordoba, Argentina\\  
 $^{3}$   Observatorio Astron\'omico, Universidad Nacional de C\'ordoba, C\'ordoba, Argentina\\  
 $^{4}$ Instituto de Astrof\'\i sica de Canarias, C/ V\'\i a L\'actea s/n, E-38205 La Laguna, Spain \\  
 $^{5}$ Departamento de Astrof\'\i sica, Universidad de La Laguna, E-38206 La Laguna, Spain \\  
$^{6}$ Physics Department, California State University East Bay, Hayward, CA 94542, USA\\  
  $^{7}$ Instituto de Astronom\'\i a y F\'\i sica del Espacio (IAFE, CONICET-UBA), Av. Inte. G\"uiraldes 2620, C1428, ZAA Buenos Aires, Argentina  
}  
\date{Accepted XXX. Received YYY; in original form ZZZ}  
  
\pubyear{2018}  
  
\begin{document}  
\label{firstpage}  
\pagerange{\pageref{firstpage}--\pageref{lastpage}}  
\maketitle  
  
\begin{abstract}  
We present new spectroscopic data of the BL Lac RGB 2243+203,  and its surroundings,    
obtained with the OSIRIS Multi Object Spectrograph (MOS) mounted in the   
Gran Telescopio Canarias (GTC). The spectra of neither the BL Lac nor its host galaxy 
show any spectral feature, thus hindering direct determination of its redshift.
The spectroscopic redshift distribution of objects in  
the MOS field of view shows four galaxies with redshift between 0.5258 and 0.5288.     
We make use of a statistical analysis to test the possibility that   
the targeted BL Lac may be a member of  that group.  
By using the spectroscopic redshifts obtained with our GTC observations,    
we found that this probability is  between 86\% and 93\%.
\end{abstract}  
  
  
  
  
\section{Introduction}  
  
At high Galactic latitudes, the $\gamma$-ray sky is dominated by 
active galactic nuclei (AGNs), especially of blazar-type.
For example,  at high energies (20 MeV - 300 GeV), 
{\it Fermi--LAT} has observed 1745 AGNs from which 
1718 are associated to either blazars, or blazar candidates~\citep[3FGL, ][]{2015Acero}.
Blazar-type sources are radio-loud AGNs,
and under the unification scheme, blazars are AGNs
in which the relativistic jets that originate close to the central massive black hole 
are oriented directly to the observer~\citep{1978Blandford}.
They consist of two kinds of objects: flat-spectrum radio quasars (FSRQs) and
BL Lac objects (BLLs).
BLLs are distinctly less luminous in
$\gamma$-rays than the FSRQs, suggesting that these two types 
trace probably different accretion regimes~\citep{2009Ghisellini,2017Ghisellini}.  
While the high luminosity of BLLs allows their detection at large distances, it also dilutes 
any spectral feature from its host
galaxy~\citep{2011Leon,2013Furniss,2014Falomo,2017Rosa}, 
 making the determination of their redshifts very unlikely.
At very high energies  (VHE; 100 GeV - 100 TeV) the spectral energy distribution (SED)   
is strongly affected by the diffuse extra--galactic background  
light (EBL) and hence appropriate corrections to recover the intrinsic  
spectral shape must be applied~\citep[e.g.][]{2008Franceschini,2011Dominguez},  
before comparing the SED  with theoretical emission  
models~\citep[e.g.][]{1993Dermer,2013Bottcher}.   
This attenuation significantly reduces the detection volume in the
Universe, which is seen in the VHE catalogue~\citep{2008Wakely}; only 78
extragalactic sources have been detected during the last few decades, 64 being
BLLs, $\sim$20\% of which have unknown or poorly-constrained redshift. Although
VHE blazars are not many, they are very important to understand the most
powerful processes in the Universe; they are candidates to be the sources of
ultra-high energy cosmic rays~\citep{2018Aab}, from which only gamma rays
and neutrinos may be observed directly~\citep{2018Ice}. They are also important
for contributing to the estimation of intergalactic magnitudes, like the EBL
~\citep{2016Ahnen,2017HESS} 
and intergalactic magnetic field~\citep{2015Caprini,2015Finke}.
In all these studies, a precise measurement of the redshift is required.  
  
In recent years several groups have been collecting data   
to obtain the redshift of the most elusive objects.    
\citet{2013Shaw}  recently compiled   
the largest sample of $\gamma$-ray selected BLLs with spectroscopic redshifts.  
They  used literature data and their own observations in 4 and 10 meter class telescopes  
resulting in successful redshift measurements for only 44\% of the 475 studied BLLs.   
In those cases where the redshift of the targets is immeasurable, 
\citet{2015Muriel} proposed a statistical method to assess the  
probability that a BLL belongs to a given system of galaxies.  
It is generally assumed that BLL nuclei are hosted by luminous elliptical  
galaxies embedded in groups of galaxies or galaxy  
clusters~\citep[e.g.][]{1990Falomo,2016Muriel,2018Torres-Zafra},  
where the nuclear activity is supposed to be enhanced by the interaction   
with a nearby companion~\citep[e.g.][]{2008Hopkins}.  
The method of~\cite{2015Muriel} (slightly modified in~\citealt{2016Rovero})  
first requires the spectroscopic identification of a galaxy group in  
the line-of-sight around the BLL; if a group is identified,   
then the probability that a given BLL is not a member of that group  
is given by  
the probability of finding a BLL in an isolated galaxy  
times the probability of finding by chance a galaxy group in the line-of-sight of the BLL.   
If that probability is low enough, the BLL is assumed to be member of the   
group and  the mean redshift of the system is assigned to it.  
The technique has been used for PKS 0447-439~\citep{2015Muriel} and  
PKS 1424+240~\citep{2016Rovero}.  
  
  
  
RGB~J2243+203 (\TS hereafter) included in the TeV catalogue~\citep{2008Wakely}   
was first reported by the 5 GHz survey carried out with the Green Bank  
telescope~\citep{1990Griffith}.   
It was classified as intermediate-energy-peaked BLL   
object in~\cite{1999Laurent}  
and as a high-synchrotron-peaked BLL galaxy in The Third {\it Fermi}-LAT  
Catalog of Sources above 10 GeV (3FHL).  
The emission of \TS  between 10 GeV and 1 TeV is better fit with  
a log--parabola function,  
\hbox{$\frac {dN}{dE}\propto(E /16.52~{\rm GeV})^{-\Gamma}$}, where  
$$\Gamma=(1.767\pm 0.256) + (0.489\pm 0.194) {\rm log}(E /16.52~{\rm GeV}),$$ 
and it has an integrated flux within this band of  (1.936$\pm$0.218)$\times$10$^{-11}$  
erg\,s$^{-1}$cm$^{-2}$~\citep{2017Ajello}.    
  
  
Sources with hard spectral index and   
detected by {\it Fermi} above 10 GeV   
are targets of interest for follow up studies at very high  
energies (VHE; E $> 100$ GeV). In fact, \TS has been detected above 160~GeV by the imaging Cherenkov telescope  
VERITAS~\citep{2017Abeysekara}, and combining their detection with different  
extragalactic background light (EBL)   
models they derived a redshift upper limit of $z<1.1$.  
  
Based on optical observations of the \TS host galaxy, and assuming that   
it can be treated as a standard candle with M$_R$=$-22.9\pm 0.5$,~\cite{2010Meisner} estimated  
a redshift lower limit of $z>0.39$.  
\cite{2013Shaw}  established a similar spectroscopic lower limit of $z>0.395$ from intervening  
absorption systems using the Low Resolution Imaging Spectrograph at the  
W. M. Keck Observatory.   
\cite{2017Paiano}  
obtained a featureless spectrum in the 4100 - 9000~\AA\, wavelength range using   
long slit observations in the Gran Telescopio Canarias (GTC)\footnote{Gran Telescopio Canarias is a   
Spanish initiative with the participation of Mexico and the US University of  
Florida, and is installed at the Roque de los Muchachos in the island of La Palma.}. 
Based also on the assumption that   
BLL host galaxies are standard candles they proposed a redshift lower limit of $z>0.22$.  
  
In this paper we complement the previous spectroscopic studies by making use of  
the OSIRIS Multi Object Spectrograph (MOS)  
mounted on the GTC. We report on the distance and environment of \ts, and  also  
use the method described by \citet{2015Muriel} to statistically associate  
a redshift to it.   
  
Throughout this paper,  
we assume a flat cold dark matter cosmology with $\Omega_{\rm M}$ =0.3  
and H$_0$=70 km s$^{-1}$ Mpc$^{-1}$.

\section{Observations and Data Reduction}  
\label{ObandRed}  
The observations were performed using the OSIRIS--MOS installed in the Nasmyth-B focus of the 10.4-m GTC  
under the program GTC5-15BMEX (PI D. Rosa Gonz\'alez).  
The observations were carried out   
in service mode, using the   
R1000R grism. The spectrum is centred at 7430~\AA\, covering the range from   
5100 to 10000 \AA\, and a sampling of 2.62~\AA/pixel. Using a fixed slit width  
of 1.2\arcsec we end up with an effective resolution measured on strong sky lines  
of around 11~\AA.  
Targets are located at different positions along the dispersion axis,  
changing the actual wavelength coverage;    
the common wavelength range covered by all spectra is 5400--9500~\AA.  
  
The total observing time was divided in two observing blocks (OB) that were   
observed on July 11$^{\rm th}$ and 28$^{\rm th}$ in 2016.  
Each OB consisted of 5 exposures of 790 seconds on target   
to facilitate removal of cosmic ray hits, and to ensure that the spectrum
of the BLL is not saturated. The OBs were accompanied by a set of ancillary files that  
included observations of Ross~640 as a  
spectro-photometric standard star, bias, flat-field and arc lamps.   
Both OBs were observed with air masses lower than 1.1 under clear gray nights   
and a seeing of between 1\arcsec and 1.2\arcsec.  
  
The data reduction was carried out using a new MOS pipeline   
described in~\citet{2016Mauricio}.  
In short, the code reduces every MOS slit by applying the usual {\small IRAF}~\footnote{IRAF is distributed by the National Optical Astronomy Observatories,
which are operated by the Association of Universities for Research
in Astronomy, Inc., under cooperative agreement with the National
Science Foundation.} scripts  
for long--slit spectra.  
To reduce the data, the three different target images were  independently corrected by bias.    
Then we stacked them to obtain a single   
spectral image where the cosmic rays were successfully removed.  
After that, every slitlet spectrum was calibrated in wavelength   
by using as a reference a He+Ne+Ar arc image.   
The dispersion solution is obtained for every single slitlet   
and in all cases  rms errors lower than 0.4~\AA\, were found.   
This uncertainty produces a systematic error in the redshift calculations of  
$\sim 10^{-4}$ at a wavelength of $\sim$5500~\AA.    
  
\begin{figure}  
	\includegraphics[width=\columnwidth]{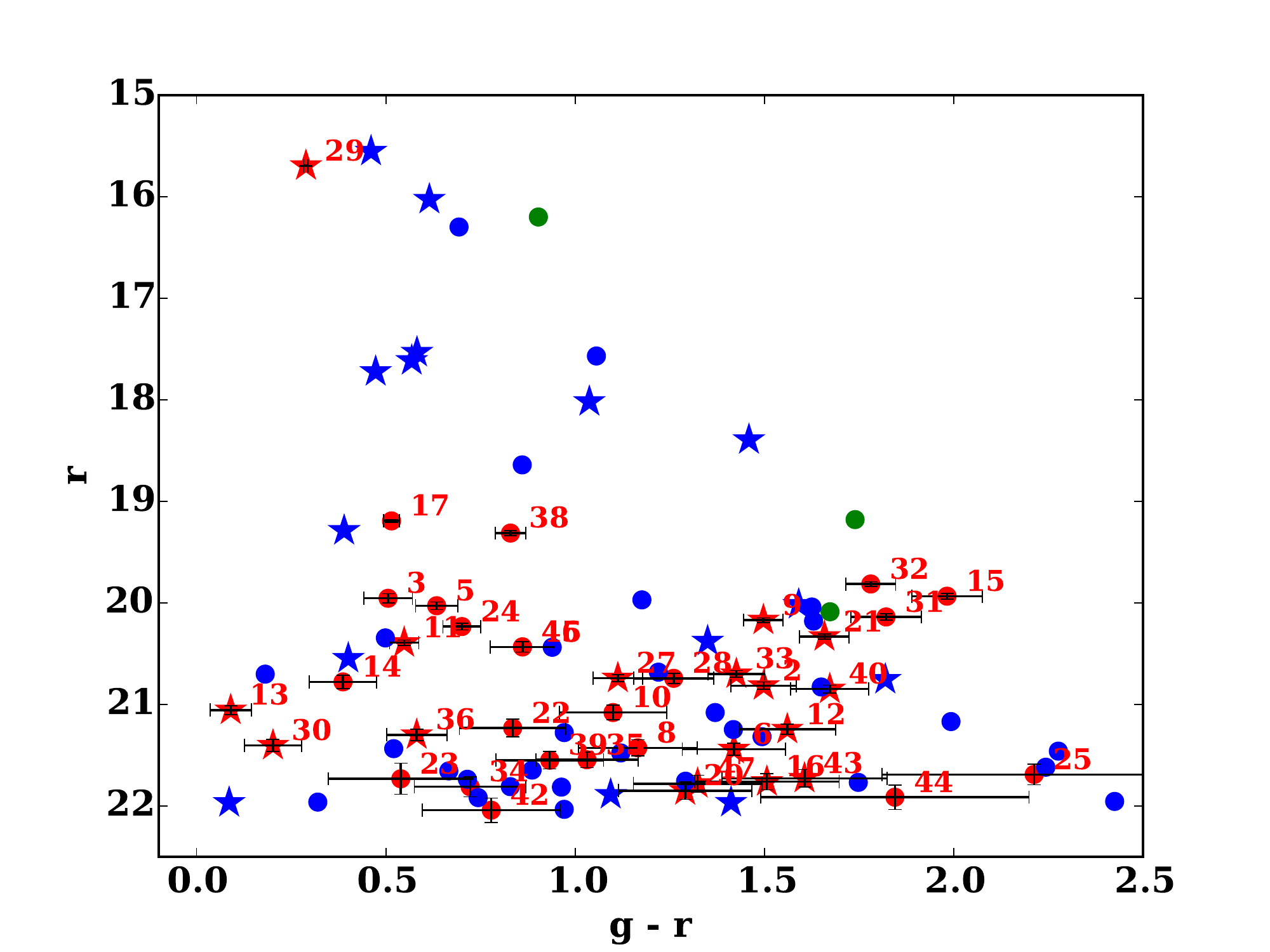}  
    \caption{  
Colour magnitude diagram for all objects brighter than $r\sim$22    
within the MOS field of view.   
One of the brightest objects corresponds to \TS (MOS-29)  
and it is located at the top left corner of the diagram.   
Objects classified as stars are  represented by star symbols, and galaxies  by  
circles.   
In red are the targets observed by us with the slit number close to   
the corresponding symbol. Green circles are objects with SDSS spectroscopy.  
In blue are the objects in the field that we did not observe.  
}  
    \label{fig:ColorMag}  
\end{figure}  
  
Finally the image is calibrated in flux by using a sensitivity curve  
obtained from the observation of Ross~640.  
The output of the pipeline is a 2-D spectral image calibrated in both    
wavelength and flux.

\begin{figure*}  
	\includegraphics[width=\textwidth]{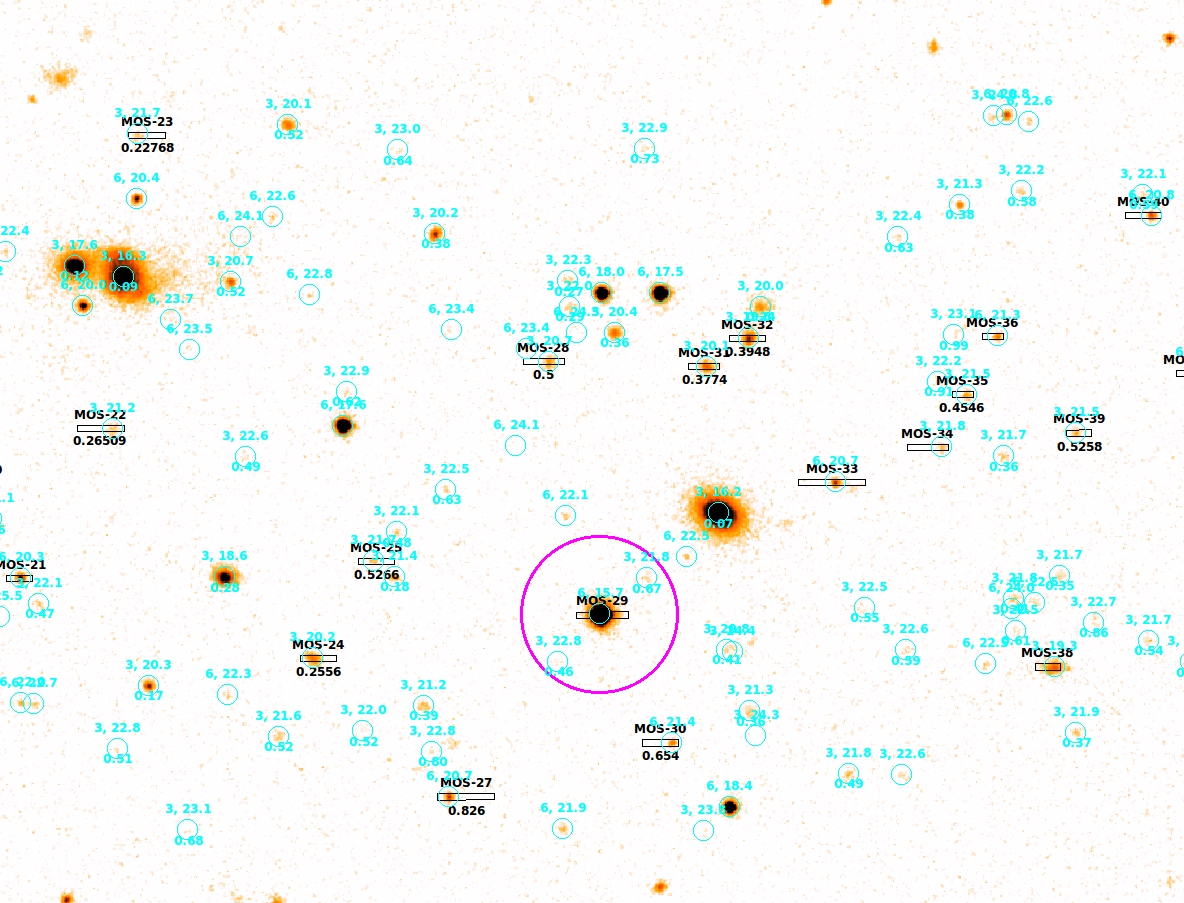}  
    \caption{$r$-band SDSS image of a FOV of  
      3.7\arcmin$\times$2.8\arcmin\,   
      showing the position of the MOS  
      slits (black rectangles).   
      Above the slits are the  slit numbers    
      and  the spectroscopic redshifts are indicated below them  (see Table~\ref{TheTable}).  
Slit 29 marks the position of \ts, and it is enclosed by a purple circle of 15\arcsec\, in radius.  
Cyan circles show the positions of SDSS objects;    
SDSS  classification (3 for galaxies, 6 for stars)   
and the $r$ magnitude are shown above each one. Below we add the photometric redshift if any.   
North is at the top, East to the left, and  the image covers partially the OSIRIS FOV.   
}  
    \label{fig:SlitPos}  
\end{figure*}

MOS allows us to obtain the spectra of several objects within an    
effective field-of-view (FOV) of around 7$^{\prime}\times2^{\prime}$.   
Excluding the stars used as astrometric guides,    
within the MOS field of view we found 92 targets brighter than $r\sim$22.  
We selected our targets based on the SDSS images centred   at the position of  
\ts. The selected FOV contains 168 SDSS objects, and   
given the nature of our program, where we do not   
know the redshift of the source, the host galaxy morphological type, or   
other extra information about the surroundings,   
we tried to cover as many objects as possible in the OSIRIS   
FOV taking care of the physical limitations of the   
mask making procedure, giving preference to those objects classified as galaxies.  
At the end, we locate the slits on top of 22 objects classified as galaxies and  
17 classified as stars. Note that some objects classified as stars  
could be distant galaxies (including \ts),    
and in fact objects classified as  
stars by SDSS, covered by slits 11, 13, 27, 30 and 43, turned out  
to be galaxies at different redshifts (Table~\ref{TheTable}).   
The colour magnitude diagram (CMD, figure~\ref{fig:ColorMag}) shows the location   
of the SDSS targets where we differentiate stars from galaxies and we mark    
those objects selected for spectroscopy.    
We cover most of the parameter space in the CMD,   
obtaining good signal to noise spectra for   
25\% of the objects brighter than $r$=22.  
  
\begin{figure*}  
	\includegraphics[width=\textwidth]{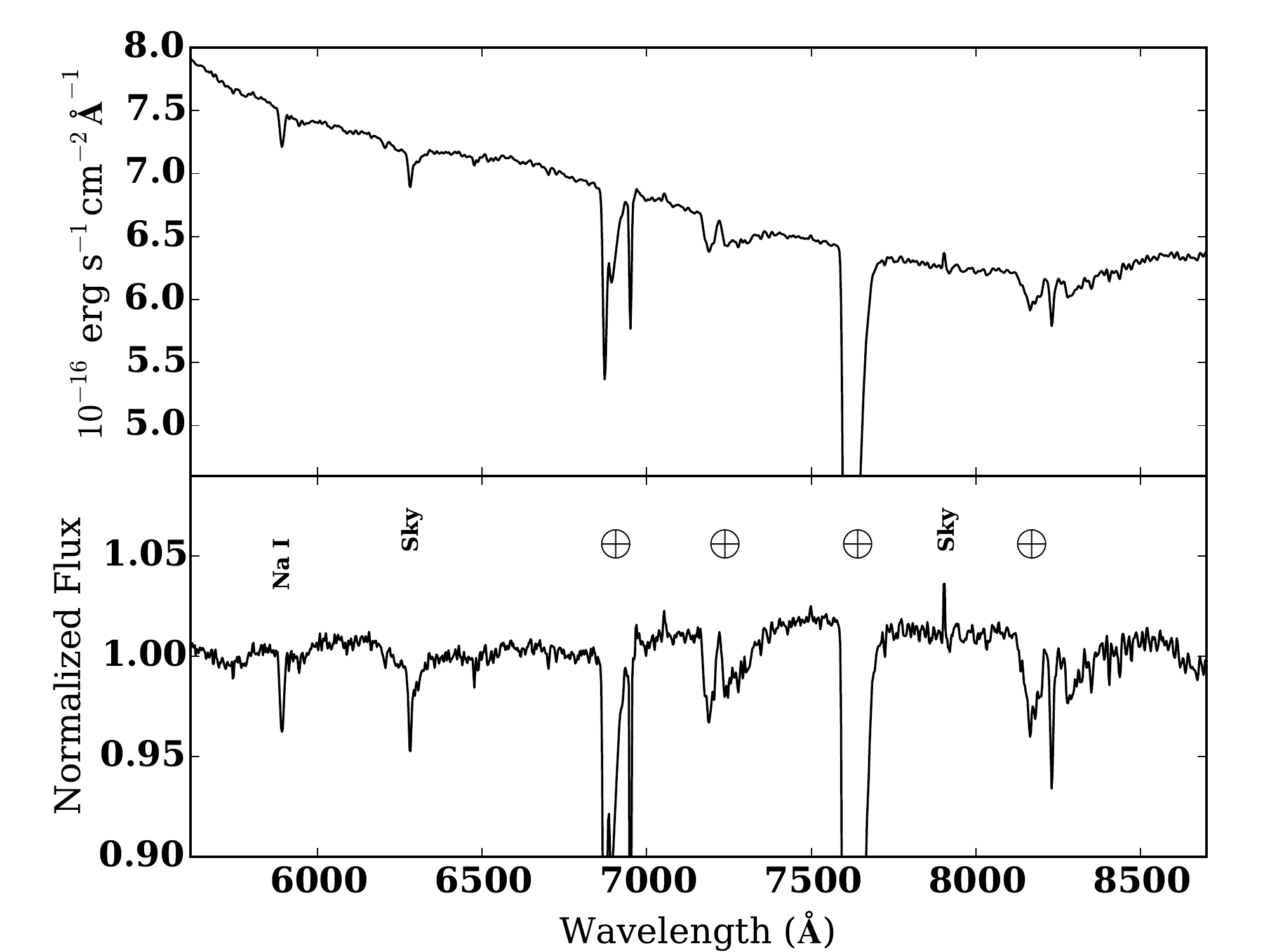}  
        \caption{The observed sky--subtracted spectrum before (top panel) and after
normalization by a smooth continuum (bottom panel). All the absorption/emission
features seen in the spectra are either due to telluric absorption lines ({\Large\earth})
or residual sky/street lamp lines ([OI]6300.304, Na\,I etc.).
Only a section of the full OSIRIS spectral coverage is plotted. The
observed bump at
8500\,\AA\ is due to the 2$^{\rm nd}$ order contribution, that could not be
properly corrected with the observations of the standard star, due to the
extremely blue spectrum of the BL Lac.
          } 
    \label{fig:HostSpec}  
\end{figure*}  
  
Figure~\ref{fig:SlitPos} shows partially   
the slit positions on top of an \hbox{$r$-band} SDSS image, where we also mark    
the fiducial stars used for astrometry and areas free of objects used for obtaining reliable sky spectra.   
The length  of the slits goes from 1\arcsec\, to 10\arcsec\, and the width is fixed at 1.2\arcsec.     
The SDSS image shows \TS in slit 29.  
  
The individual spectrum for each object was extracted from the calibrated  
 2-D spectral image by using the {\small IRAF} task $apall$.   
In most of the cases the continuum is   
well detected and a 4$^{\rm th}$ order polynomial function was good enough to fit   
the trace along the dispersion axis.   
In the cases where the continuum was not well detected   
we used  the trace solutions found for a nearby object as reference.   
  
In general the extraction window is centred on the peak   
of the continuum, however for the case of the extraction   
of the spectrum corresponding to \TS (MOS-29), and   
keeping in mind that the emission coming from the   
centre is expected to be featureless, we choose two apertures that avoid the  
central pixels. Details of these cuts are given in the next section.  
  
Once the 1-D individual spectra were extracted, we located by eye different   
spectral features. We looked for the most common lines observed in extra--galactic sources,   
both in absorption (e.g. Ca~\textsc{II} H\&K, Mg band, NaD) and in emission (e.g. hydrogen   
recombination lines, [OII]$\lambda$3727, [OIII]$\lambda$5007, [NII]$\lambda$6583).  
Once the lines were identified in a given spectrum we fit a Gaussian profile   
which returns the position of the peaks   
(we used the {\small IRAF}/$splot$ command stroke $k$).   
The peak position  was used to estimate the   
corresponding redshifts for each individual line. The final redshift  
is the median of all measured  values. The error is given   
by the standard deviation when we have more than two  
lines in the spectrum, or the difference   
between the two redshift values    
when we have only  two lines identified in the spectrum.   
The systematic error in redshift of $10^{-4}$\,   
due to uncertainties in the   
wavelength calibration described in Section~\ref{ObandRed}   
was added in quadrature.   
The spectroscopic redshifts and the errors are included in Table~\ref{TheTable}.  
We also included the number of lines used to estimate the redshifts and   
whether they were found in emission or in absorption.  
MOS-13, an object classified as a star by the SDSS, is the only target where  
only one line at $\sim6042$~\AA\ was identified. Its redshift was calculated assuming that the observed broad emission line corresponds   
to   Mg{\,\small II\,}$\lambda$2797.

\section{Results}  
\label{Results} 
\subsection{Spectrum of \TS}  
  
The flux calibrated spectrum of the \TS galaxy is plotted in  
Figure~\ref{fig:HostSpec} (Top).
The spectrum was barely smoothed by using a  Savitzky--Golay filter
\citep[e.g.][]{1989Press}, 
with a window size of 3 pixels and a one-degree polynomial. 
\footnote{The Savitzky--Golay filter is a particular type of low-pass filter. We make
use of the routine provided by the SciPy organization at SciPy.org. Infor-
mation of the algorithm and multiple references can be found in the SciPy
pages.}

The spectrum corresponds to the entire BLL spectrum including  
the central pixels where the featureless continuum has its maximum  
contribution.  
It was extracted using an aperture of 8 pixels (2\arcsec), and it has   
a signal to noise ratio (S/N) of $\sim$130 which is about two times  
worse than the one published by~\cite{2013Shaw} based on  
observations with the Low Resolution Imaging Spectrograph at the W. M. Keck Observatory,   
and within  the range of S/N values presented recently by~\cite{2017Paiano}   
obtained with the GTC.
We also show the position of  
the sodium line and several telluric absorption lines.

In our previous work  where we examined the spectrum of the BLL   
3FGL J0909.0+2310, we noticed that avoiding the central part of the continuum  
emission facilitates the detection of absorption spectral features with better  
signal to noise ratio~\citep{2017Rosa}.  
Following this result we realized two different cuts which avoid the featureless continuum.
One corresponds to an extraction window centred 2 pixels to the  
west of the continuum peak,  and the other to a centre located 2 pixels to the east.  
Both cuts were performed using an extraction window of 4 pixels (1\arcsec).  
We did not find any spectral feature in any of the three spectra analyzed.  
  
  
\begin{figure}  
	\includegraphics[width=\columnwidth]{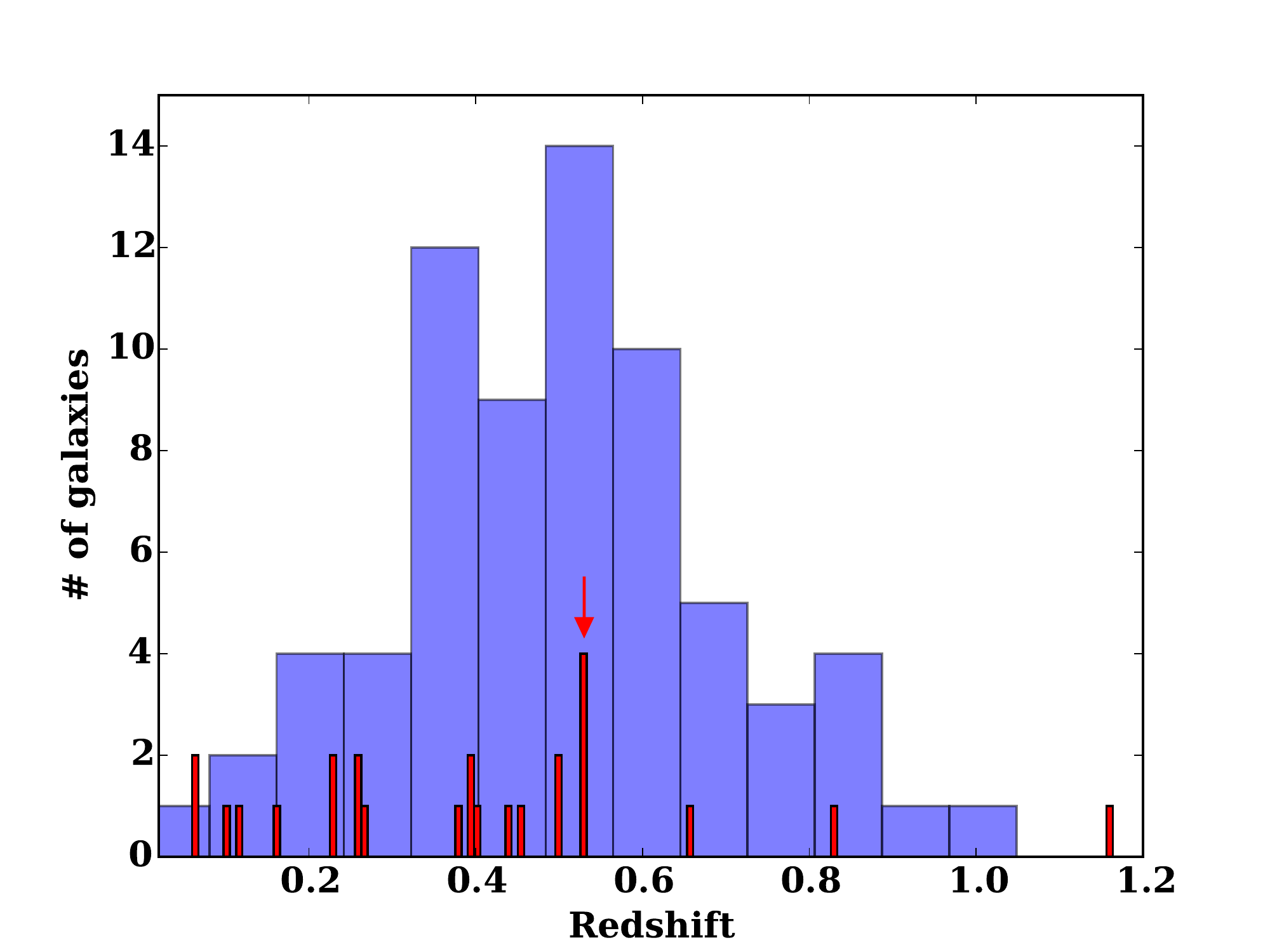}  
    \caption{Spectroscopic  (red)   
and photometric (blue) redshift  distributions   
of the targets within the OSIRIS FOV.  
The bin size of the spectroscopic distribution is  0.0075, and
0.08 for the photometric distribution.
One small group of galaxies with spectroscopic redshifts around 0.53   
is identified and highlighted with the red arrow.   
The  red histogram includes the three objects with SDSS spectroscopy.}  
    \label{fig:Zdist}  
\end{figure}

\subsection{Redshift of Galaxies in the OSIRIS FOV}  
\label{ZofMain}  
We obtain spectroscopic redshifts for 22 of the 39 observed targets.   
Targets for which we cannot extract the redshift include   
objects close to the slit borders and those where   
the signal to noise ratio was too low to obtain any spectral feature.    
Table~\ref{TheTable} lists the SDSS properties of the selected targets,    
and the spectroscopic redshifts.  
  
We notice that, additionally to our observations, three galaxies within the observed FOV have SDSS  
spectroscopy and they are included in the data release 12~\citep{2015Alam}.    
We include in table~\ref{SDSSzetas} the  
position and spectroscopic redshifts of these sources.

To study the existence of galaxy groups within the OSIRIS FOV    
we create a histogram showing   
the redshift distribution based on photometric and   
spectroscopic redshifts (figure~\ref{fig:Zdist}).  
The spectroscopic data come from our MOS observations   
and the 3 galaxies from SDSS.  
The bin of the  spectroscopic redshift histogram is of 0.0075, and corresponds to a velocity of around  
2000\kms. Values this large have been found in the study of the velocity dispersion  
of massive galaxy clusters and is ideal for our blind search of  
groups.  
The photometric redshifts were obtained from   
the SDSS data release 12~\citep{2016Beck} which includes all the   
galaxy--type objects within the OSIRIS FOV  (7.4$^{\prime}\times 2.3^{\prime}$ centered in \ts).    
The bin size of the photometric redshift distribution was fixed in 0.08 
which corresponds to the median of the photometric redshift errors of the selected galaxies.

\begin{figure}  
	\includegraphics[width=\columnwidth]{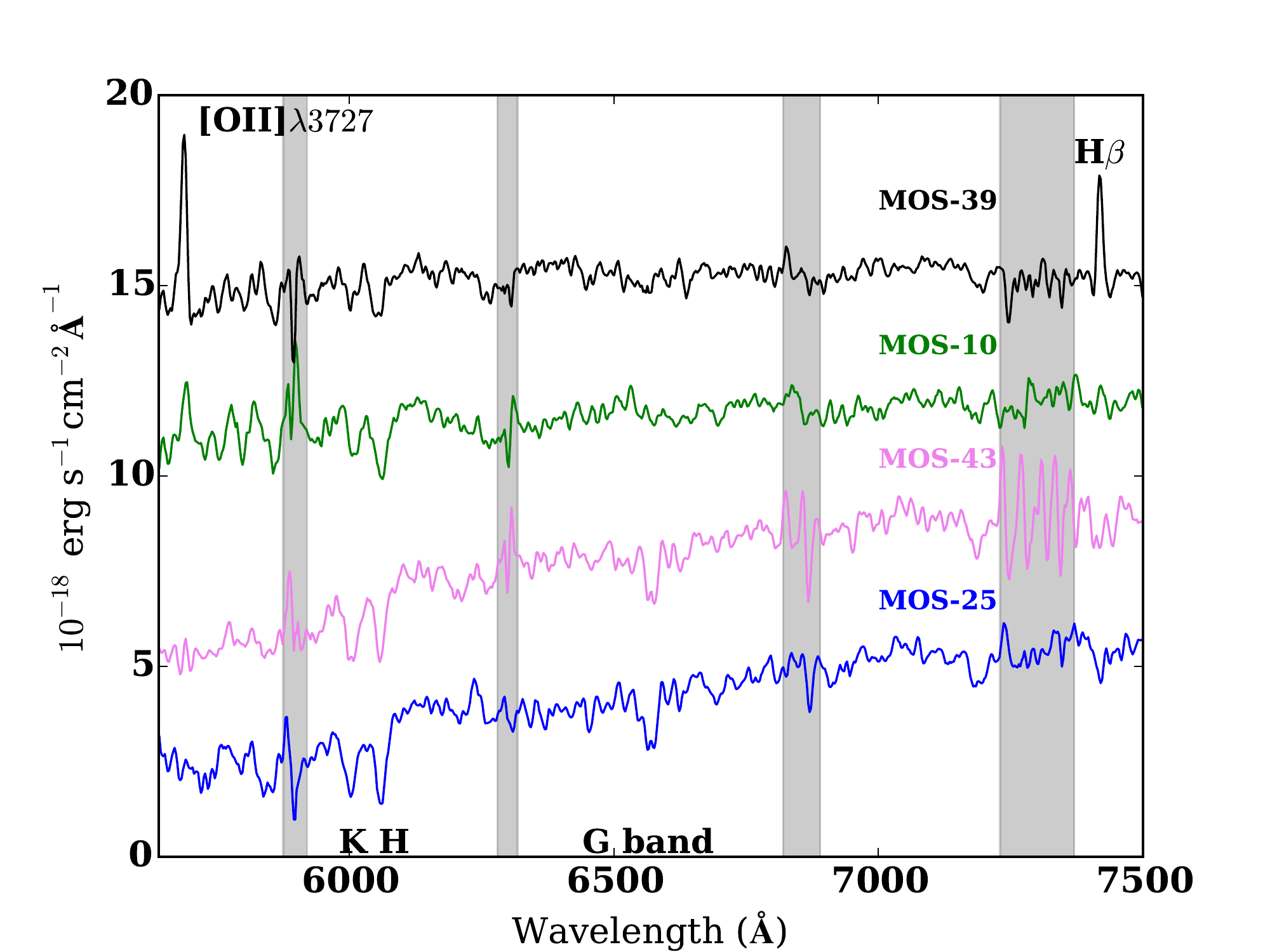}  
        \caption{GTC spectra of the galaxies with spectroscopic redshift at  
          redshift around 0.53. The position of common absorption and emission  
          lines are marked. The gray vertical bands show the position of strong  
          sky lines where the sky substraction is far from perfect and fake  
          structures appear in the spectra.             
          From the bottom to the top the spectra have been artificially  
          shifted by adding constant continuum values of 4, 8 and 12$\times  
          10^{-18}$\ergscm \AA$^{-1}$.  The spectra were filtered by using
          a  Savitzky--Golay filter with a window size of 3 pixels and a
          polynomial of order one.
        }  
    \label{fig:GroupSpct}  
\end{figure}

A small peak with four members,   
appears in the spectroscopic distributions at redshift  0.53.   
As mentioned in the introduction  BLL are believed to be hosted by elliptical
galaxies~\citep{2016Muriel} and  are primarily found in  groups of three or more members with typical virial
radius of around 1 Mpc~\citep{2002Merchan}. 
In fact, three of the four  
galaxies with redshift around 0.53 show the typical absorption lines  
observed in local elliptical galaxies (figure~\ref{fig:GroupSpct}).  
The possibility that \TS belongs to that group at $z\sim$0.53 is explored further in the  
next section.   

\subsection{Statistical Redshift }  
  
\begin{figure}  
	\includegraphics[width=\columnwidth]{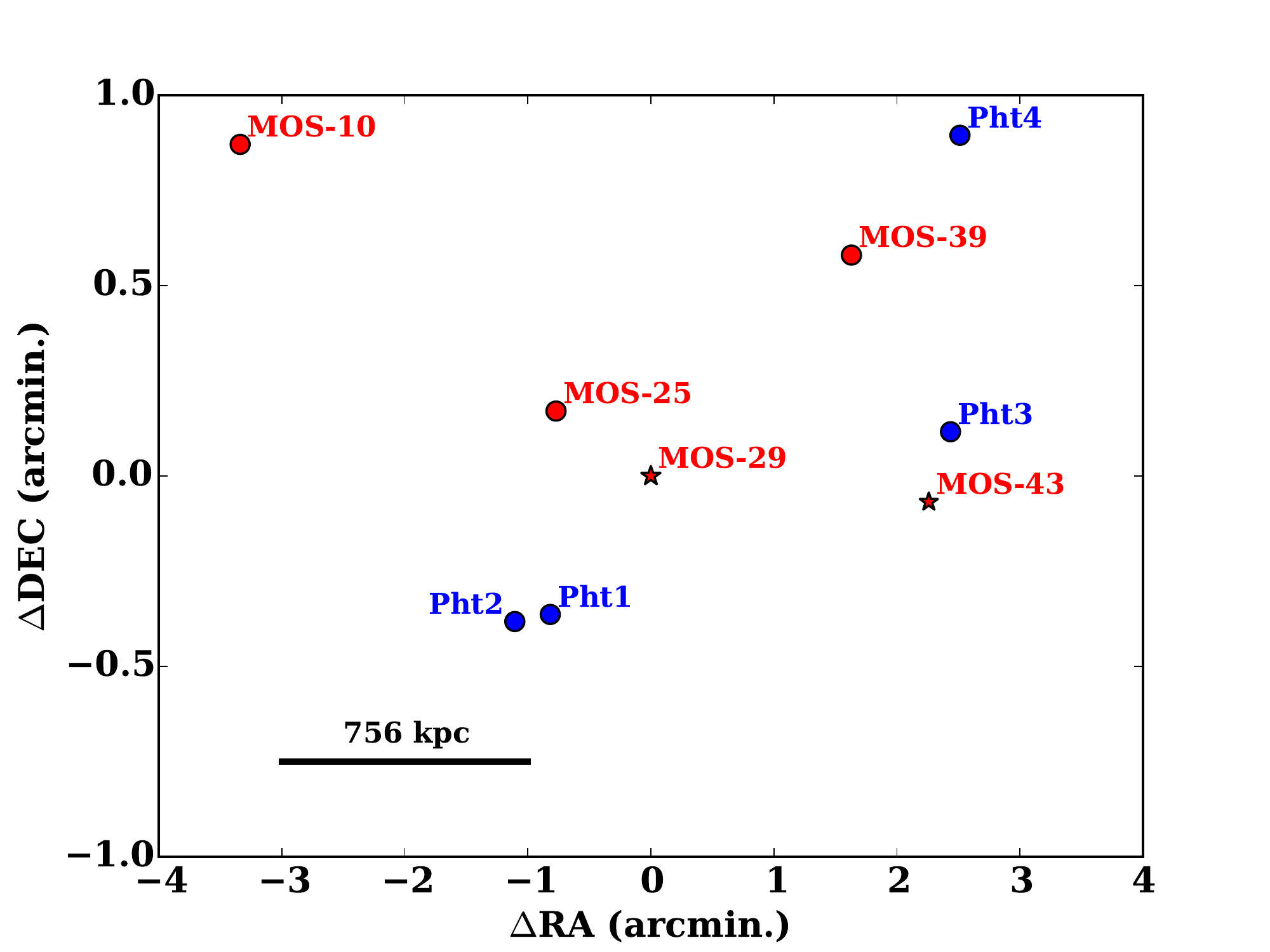}  
        \caption{Relative position of the galaxies with redshits around 0.53.  
        The diagram is centred in the position of MOSS--29 (\ts) and we plotted in  
        red the galaxies with spectroscopic redshift and in blue the ones with  
        photometric redshifts. Both MOS--29 and MOS--43 were classified as stars  
        by the SDSS, and we use a star--symbol to distinguish them.   
        The physical scale is displayed in the bottom corner of the image.    
        }  
    \label{fig:RelPos}  
\end{figure}

As seen in the previous section, the redshift distribution (figure~\ref{fig:Zdist}) in the line  
of sight of \TS shows the presence of a group of galaxies with 4  
spectroscopic members. It is also observed that the group  
could be more numerous if we take into account the four galaxies with  
photometric redshift at the same distance.  
The relative position of these two groups of galaxies with respect to \TS are  
plotted in figure~\ref{fig:RelPos}, and different properties of them are presented  
in tables~\ref{TheTable}, and~\ref{PhotZ}.  

Using only the four galaxies with spectroscopic redshift, 
we computed the mean  
redshift ($z_{mean}$) and  the velocity dispersion ($\sigma_{v}$)  
using the Gapper estimator defined by \cite{1990Beers} and   
the virial radius ($R_{vir}$) following~\cite{2013Nurmi}.  
We obtained the following  
values:  $z_{mean} = 0.527 \pm 0.001$; $R_{vir}$ = 0.84 Mpc (2.2\arcmin\, at that  
distance) and  
$\sigma_{v} = 197$ km s$^{-1} \pm 124$.
Given that the mean redshift and the velocity dispersion are computed using only
four galaxies, the quoted values are susceptible to errors associated with the statistics of
small numbers and should be taken with caution.

  
We follow the procedure described in \citet{2015Muriel} and \citet{2016Rovero}  
to estimate the probability that \TS is a member of the  
group of galaxies at $z=0.527$ described by the parameters calculated previously.   
In short the method can be described in the following steps,  
i) we use the most complete inner field of the zCOSMOS 20k  
group catalog \citep[][hereafter 20k catalog]{2012Knobel},  
which has 1498 identified groups; ii) we test the completeness of our sample  
of spectroscopic redshifts computing the fraction of galaxies per bin  
of apparent magnitude that have spectroscopic redshift; iii) we  
extract from the 20k catalog a random subsample of galaxies (a pruned  
catalog) by selecting galaxies that have approximately the same  
magnitude distribution as our spectroscopic sample. In this process we
take into account that the mean spectroscopic completeness of   
zCOSMOS is 56\%; iv) the pruned catalog of galaxies is used to define  
a catalog of groups that includes those systems that survive   
the random selection of galaxies (the new richness of groups is also  
computed); v) within the pruned catalog of groups of galaxies, we  
select random positions and see the coincidence with groups with four  
or more members; vi) we repeat the whole procedure 100 times and  
compute the mean probability to find by chance a group of galaxies  
like the observed one. We found a probability of $22.5 \pm 2.0$\% of  
finding a group of four or more members in an observation similar to  
the one reported here. The error corresponds to the standard deviation  
of the 100 realizations.  
  
  
If the group of galaxies found in the line of sight does not host the  
BLL, it means that the host galaxy is an isolated object. As it was  
pointed out, \cite{2016Muriel} found that at least $67\pm 8$\% of the BLLs  
are in galaxies that are members of groups or clusters of galaxies.  
Since the probability of the BLL to be hosted by an isolated galaxy is  
independent of the probability computed above of finding a group by  
chance, then the probability of \TS not being associated with the  
group of four members is the joint probability of having both, a group  
by chance ($22.5 \pm 2.0$\%) and an isolated BLL ($\lesssim 33 \pm8$\%)  
is $\lesssim 7.4 \pm 1.8$\% (the error is computed using  
propagation of uncertainties).
This implies that -- if we assumed that the BL Lac is an isolated galaxy
-- the possible membership of \TS to the group of galaxies found at z = 0.527  
has an occurrence probability of $\sim 92.6 \pm 1.8$\%.
On the other hand the joint probability  of finding a group with five members
  and the BL Lac belonging to an undetected group of galaxies is of 14$\pm 2$\%,
  and therefore the probability of \TS being a member of the group at  z = 0.527
  would be of 86$\pm 2$\%.

  
\section{Conclusions}  
  
We have presented high signal to noise optical spectrum of the elusive object \ts.  
The spectrum does not show any spectral  
features that can be used to determine its distance with precision.  
Based on the spectroscopic observations of the  
BLL surroundings we found four galaxies at a   
redshift of around 0.53. Three of them have the typical absorption lines  
studied in local elliptical galaxies.  
By using a statistical analysis where we take into  
account that BLL objects are usually found in galaxy groups   
we found that, if we use only the spectroscopic confirmed object at redshift  
$\sim$0.53, then the probability that \TS would be a member of that group is
between 86\% and 93$\%$.
Based on these results we propose that RGB 2243+203
is hosted by a galaxy for which we estimate an statistical redshift of $z=0.527\pm0.001$.
Being a member of that group is in agreement with the  
redshift upper limits proposed by the EBL studies reported by   
the VERITAS team~\cite{2017Abeysekara} and the lower limits obtained  
by recent optical observations~\citep{2013Shaw,2017Paiano}.   
  
\begin{table*}  
	\centering  
	\caption{Targets observed with  GTC-MOS. Columns   1 to 6 give:   
slit number (ID), SDSS coordinates,  type (3 for galaxies, 6 for stars), SDSS $r$ band, and SDSS $g-r$ colour.   
The  spectroscopic redshift ($z_{\rm sp}$), its error and the number of lines (N) used for the  
redshift determination are given in columns 7 and 8.   
Column 8 also indicates whether the lines were observed in absorption (A) or in  
emission (E). Column 9 indicates the projected and physical distance to \ts,  
for those galaxies with spectroscopic redshift around 0.53.   
The slits on  fiducial stars or those used for sky measurements are not shown.  
}  
	\label{TheTable}  
	\begin{tabular}{lccccccrcc} 
		\hline  
ID & RA($^{\rm o}$) & DEC($^{\rm o}$)& Type & $r$\,(mag.) & $g-r$\,(mag.) & $z_{\rm sp}$\  \  \ \ \ & N  & Distances & Comments \\      
\ & \ & \ & \ & \ & \ & \ \ & \  & \small (arcmin. \& Mpc) & \ \\      
\hline  
MOS-2 & 341.05611 & 20.35854 & 6 & 20.81$\pm$0.04 & 1.50$\pm$0.09 &  -- & -- & --   & (b) \\   
MOS-3 & 341.05403 & 20.36081 & 3 & 19.95$\pm$0.05 & 0.51$\pm$0.06 &  0.0625$\pm$0.0002 & 5/E & --   &   \\   
MOS-5 & 341.04905 & 20.33917 & 3 & 20.03$\pm$0.04 & 0.63$\pm$0.06 &  0.1137$\pm$0.0001 & 4/E & --   &   \\   
MOS-6 & 341.04592 & 20.35365 & 6 & 21.44$\pm$0.06 & 1.42$\pm$0.14 &  -- & -- & --   & (b) \\   
MOS-8 & 341.03945 & 20.36384 & 3 & 21.43$\pm$0.08 & 1.17$\pm$0.16 &  0.2272$\pm$0.0001 & 4/E & --   &   \\   
MOS-9 & 341.03655 & 20.35353 & 6 & 20.17$\pm$0.02 & 1.50$\pm$0.05 &  -- & -- & --   & (b) \\   
MOS-10 & 341.03362 & 20.36537 & 3 & 21.08$\pm$0.07 & 1.10$\pm$0.14 &  0.5280$\pm$0.0010 & 3/A &  3.4\hskip 0.5cm  1.3  & (e) \\   
MOS-11 & 341.03178 & 20.34478 & 6 & 20.39$\pm$0.03 & 0.55$\pm$0.04 &  0.4410$\pm$0.0020 & 4/A & --   &   \\   
MOS-12 & 341.02989 & 20.35458 & 6 & 21.24$\pm$0.05 & 1.56$\pm$0.13 &  -- & -- & --   & (b) \\   
MOS-13 & 341.02779 & 20.37159 & 6 & 21.05$\pm$0.04 & 0.09$\pm$0.05 &  1.1590 & 1/E & --   & (d) \\   
MOS-14 & 341.02600 & 20.37331 & 3 & 20.78$\pm$0.07 & 0.39$\pm$0.09 &  0.0985$\pm$0.0001 & 4/E & --   &   \\   
MOS-15 & 341.02449 & 20.34902 & 3 & 19.93$\pm$0.03 & 1.98$\pm$0.09 &  0.4000$\pm$0.0008 & 6/A & --   &   \\   
MOS-16 & 341.02261 & 20.34985 & 6 & 21.76$\pm$0.08 & 1.51$\pm$0.19 &  -- & -- & --   & (b) \\   
MOS-17 & 341.02033 & 20.35083 & 3 & 19.19$\pm$0.01 & 0.51$\pm$0.02 &  0.1600$\pm$0.0100 & 4/E & --   &   \\   
MOS-20 & 341.01367 & 20.35787 & 6 & 21.85$\pm$0.09 & 1.29$\pm$0.18 &  -- & -- & --   & (b) \\   
MOS-21 & 341.01118 & 20.35279 & 6 & 20.33$\pm$0.03 & 1.66$\pm$0.07 &  -- & -- & --   & (b) \\   
MOS-22 & 341.00660 & 20.36087 & 3 & 21.23$\pm$0.09 & 0.83$\pm$0.14 &  0.2651$\pm$0.0001 & 4/E & --   &   \\   
MOS-23 & 341.00392 & 20.37651 & 3 & 21.73$\pm$0.15 & 0.54$\pm$0.19 &  0.2277$\pm$0.0001 & 4/E & --   &   \\   
MOS-24 & 340.99415 & 20.34854 & 3 & 20.23$\pm$0.03 & 0.70$\pm$0.05 &  0.2556$\pm$0.0001 & 8/E & --   &   \\   
MOS-25 & 340.99083 & 20.35370 & 3 & 21.69$\pm$0.10 & 2.21$\pm$0.40 &  0.5266$\pm$0.0007 & 5/A &  0.8\hskip 0.5cm  0.3  & (e) \\   
MOS-27 & 340.98573 & 20.34112 & 6 & 20.74$\pm$0.04 & 1.11$\pm$0.07 &  0.8260$\pm$0.0020 & 3/A & --   &   \\   
MOS-28 & 340.98131 & 20.36440 & 3 & 20.74$\pm$0.05 & 1.26$\pm$0.11 &  0.5000$\pm$0.0004 & 2/A & --   &   \\   
MOS-29 & 340.97798 & 20.35086 & 6 & 15.69$\pm$0.00 & 0.29$\pm$0.01 &  -- & -- & --   & (a) \\   
MOS-30 & 340.97466 & 20.34401 & 6 & 21.40$\pm$0.06 & 0.20$\pm$0.08 &  0.6540$\pm$0.0020 & 2/A & --   &   \\   
MOS-31 & 340.97217 & 20.36413 & 3 & 20.14$\pm$0.03 & 1.82$\pm$0.09 &  0.3774$\pm$0.0003 & 6/A & --   &   \\   
MOS-32 & 340.96972 & 20.36563 & 3 & 19.81$\pm$0.02 & 1.78$\pm$0.07 &  0.3948$\pm$0.0005 & 6/A & --   &   \\   
MOS-33 & 340.96487 & 20.35790 & 6 & 20.70$\pm$0.03 & 1.43$\pm$0.08 &  -- & -- & --   & (b) \\   
MOS-34 & 340.95978 & 20.35983 & 3 & 21.81$\pm$0.10 & 0.72$\pm$0.15 &  -- & -- & --   & (c) \\   
MOS-35 & 340.95752 & 20.36262 & 3 & 21.54$\pm$0.08 & 1.03$\pm$0.13 &  0.4546$\pm$0.0002 & 3/E & --   &   \\   
MOS-36 & 340.95587 & 20.36576 & 6 & 21.30$\pm$0.06 & 0.58$\pm$0.08 &  -- & -- & --   & (c) \\   
MOS-38 & 340.95286 & 20.34806 & 3 & 19.31$\pm$0.02 & 0.83$\pm$0.04 &  -- & -- & --   & (c) \\   
MOS-39 & 340.95081 & 20.36053 & 3 & 21.55$\pm$0.09 & 0.93$\pm$0.14 &  0.5258$\pm$0.0007 & 3/E &  1.7\hskip 0.5cm  0.7  & (e) \\   
MOS-40 & 340.94721 & 20.37217 & 6 & 20.85$\pm$0.04 & 1.67$\pm$0.10 &  -- & -- & --   & (b) \\   
MOS-42 & 340.94307 & 20.36790 & 3 & 22.04$\pm$0.12 & 0.78$\pm$0.18 &  -- & -- & --   & (c) \\   
MOS-43 & 340.94033 & 20.34972 & 6 & 21.73$\pm$0.09 & 1.61$\pm$0.22 &  0.5268$\pm$0.0002 & 3/A &  2.3\hskip 0.5cm  0.9  & (e) \\   
MOS-44 & 340.93744 & 20.35147 & 3 & 21.91$\pm$0.12 & 1.84$\pm$0.35 &  -- & -- & --   & (c) \\   
MOS-45 & 340.93381 & 20.37821 & 3 & 20.43$\pm$0.05 & 0.86$\pm$0.09 &  0.2573$\pm$0.0006 & 4/E & --   &   \\   
MOS-46 & 340.93113 & 20.37710 & 3 & 20.43$\pm$0.05 & 0.86$\pm$0.09 &  -- & -- & --   & (c) \\   
MOS-47 & 340.92678 & 20.37309 & 6 & 21.78$\pm$0.08 & 1.32$\pm$0.17 &  -- & -- & --   & (b) \\   
		\hline  
	\end{tabular}  
\\  
(a) Main target, (b)  field star,  (c) no spectral features detected, (d)  
the reported $z=$1.1590 assume that the detected line correspond to Mg{\,\small  
  II\,}$\lambda$2797, and (e) galaxies with spectroscopic redshift around 0.53.  
\end{table*}

\begin{table}  
	\centering  
	\caption{  
Objects within the FOV and SDSS spectroscopy.   
}  
	\label{SDSSzetas}  
	\begin{tabular}{ccc} 
		\hline  
RA($^{\rm o}$) & DEC($^{\rm o}$)  & $z_{\rm sp}$ \\      
\hline  
340.93366 & 20.34935 & 0.37943$\pm$0.00008 \\  
340.97133 & 20.35645 & 0.06514$\pm$0.00001 \\  
340.99589 & 20.37720 & 0.50023$\pm$0.00011  \\  
		\hline  
	\end{tabular}  
\\  
\end{table}

\begin{table*}  
	\centering  
	\caption{  
Properties of the SDSS galaxies with photometric redshift around 0.53. The  
columns show the position, r-magnitude, colors, photometric redshift and  
distances to \ts.  
        }  
        \label{PhotZ}  
	\begin{tabular}{cclccccc}	\hline     
ID & RA($^{\rm o}$) & DEC($^{\rm o}$)&  $r$\,(mag.) & $g-r$\,(mag.) & $z_{\rm  ph}$ & Distance \\   
&  & &  &  &  & \small (arcmin. \& Mpc) \\ 	\hline  
Pht-1 & 340.99161 &  20.344800 & 21.96$\pm$0.21 & 0.32$\pm$0.26 & 0.520192 $\pm$ 0.191746 & 0.9\hskip 0.5cm  0.3\\   
Pht-2 & 340.99641 &  20.344489 & 21.65$\pm$0.10 & 0.89$\pm$0.17 & 0.520293 $\pm$ 0.137049 & 1.2\hskip 0.5cm  0.4\\   
Pht-3 & 340.93739 &  20.352797 & 24.14$\pm$0.62 & -1.03$\pm$0.65 & 0.523156$\pm$ 0.217098 & 2.4\hskip 0.5cm  0.9\\   
Pht-4 & 340.93611 &  20.365772 & 22.26$\pm$0.15 & 1.22$\pm$0.30 & 0.519344 $\pm$ 0.062896 & 2.7\hskip 0.5cm  1.0\\   
\hline  
	\end{tabular}  
\\  
\end{table*}  
   
\section*{Acknowledgements}  
  
We thank the support team at GTC  and the anonymous referee for helpful suggestions.
Based on observations made with the Gran Telescopio Canarias (GTC), installed at  
the Spanish Observatorio del  
Roque de los Muchachos of the Instituto de Astrof\'\i sica de Canarias, in the island of La Palma.  
This work is partly financed   
by CONACyT -- the  
Mexican research Council --  research grants CB-2010-01-155142-G3  
(PI:YDM) and CB-2011-01-167281-F3 (PI:DRG). SCL thanks  
CONACyT for her studentship.   
This paper has been partially supported with grants from CONICET -- the  
Argentinian research Council -- and SeCyT,  
Universidad Nacional de C\'ordoba, Argentina.  
  
    Funding for the SDSS and SDSS-II has been provided by the Alfred P. Sloan Foundation, the Participating Institutions, the National Science Foundation, the U.S. Department of Energy, the National Aeronautics and Space Administration, the Japanese Monbukagakusho, the Max Planck Society, and the Higher Education Funding Council for England. The SDSS Web Site is http://www.sdss.org/.  
  
    The SDSS is managed by the Astrophysical Research Consortium for the Participating Institutions. The Participating Institutions are the American Museum of Natural History, Astrophysical Institute Potsdam, University of Basel, University of Cambridge, Case Western Reserve University, University of Chicago, Drexel University, Fermilab, the Institute for Advanced Study, the Japan Participation Group, Johns Hopkins University, the Joint Institute for Nuclear Astrophysics, the Kavli Institute for Particle Astrophysics and Cosmology, the Korean Scientist Group, the Chinese Academy of Sciences (LAMOST), Los Alamos National Laboratory, the Max-Planck-Institute for Astronomy (MPIA), the Max-Planck-Institute for Astrophysics (MPA), New Mexico State University, Ohio State University, University of Pittsburgh, University of Portsmouth, Princeton University, the United States Naval Observatory, and the University of Washington.

  
  
  


  
  
  
  
\bsp	
\label{lastpage}  
\end{document}